\documentclass[letterpaper]{article}

\usepackage{natbib,alifeconf}  
\usepackage{booktabs}
\usepackage{amsmath}
\usepackage{adjustbox} 
\usepackage{listings}
\usepackage{algorithm}
\usepackage{algpseudocode}

\usepackage{url,hyperref,cleveref}
\title{ALIFE2024 template}

\title{Emergent Collective Reproduction via Evolving Neuronal Flocks}

\author{
    Nam H. Le$^{1, 4}$\thanks{Corresponding Author: Nam H. Le. Email: namlehai90@gmail.com. Formerly at the University of Southampton, current position at the University of Vermont.},
    Richard A. Watson$^{1}$, \and
    Chris Buckley$^2$ \\
    Michael Levin$^3$ \\
    \mbox{}\\
    $^1$University of Southampton, UK \\
    $^2$Uiversity of Sussex, UK \\
    $^3$Tufts University, USA \\
    $^4$University of Vermont, USA \\
} 

\begin{document}

\maketitle

\begin{abstract}

This study facilitates the understanding of evolutionary transitions in individuality (ETIs) through a novel artificial life framework, named VitaNova, that intricately merges self-organization and natural selection to simulate the emergence of complex, reproductive groups. By dynamically modelling individual agents within an environment that challenges them with predators and spatial constraints, VitaNova elucidates the mechanisms by which simple agents evolve into cohesive units exhibiting collective reproduction. The findings underscore the synergy between self-organized behaviours and adaptive evolutionary strategies as fundamental drivers of ETIs. This approach not only contributes to a deeper understanding of higher-order biological individuality but also sets a new precedent in the empirical investigation of ETIs, challenging and extending current theoretical frameworks.
   
\end{abstract}

\section{1. Introductory Background}

Understanding the mechanisms behind mysterious evolutionary transitions in individuality (ETIs) is a central narrative in contemporary biology \cite{okasha2005multilevel, szathmary2015toward}. These transitions, which encompass the evolutionary milestones enabling discrete biological entities to coalesce into complex, higher-order wholes, pose profound questions about the origins of collective reproduction and complex life forms. At the heart of understanding ETIs lies the exploration of how new levels of biological organisation emerge and the dynamics by which these levels attain and sustain the capability for collective reproduction \cite{smith1997major}.

ETI research examines the complex interplay between individual fitness and collective fitness in emergent groups. Historically, studies have fluctuated between focusing on the reproductive interests of the components (e.g., cells or individuals) and the overall fitness of the whole (e.g., multicellular organisms or social groups) \cite{michod2003reorganization}. This body of work aims to elucidate the conditions under which evolutionary pressures facilitate a ``decoupling,'' where selective pressures on individual components are intricately linked but distinct from those on the collective entity \cite{michod2000darwinian}.

Moreover, traditional approaches have often leaned on theoretical models that, while invaluable, lack the dynamic representation of evolutionary processes and the spontaneous emergence of higher-level individuality \cite{stewart1997evolutionary, heylighen2000evolutionary}. This fragmentation in ETI research highlights the need for innovative methodologies capable of integrating these complex dynamics into a cohesive understanding. Artificial Life (ALife) simulations emerge as a promising frontier in this context, offering a novel lens through which the spontaneous emergence of collective behaviours and higher-order individuality can be observed and analysed \cite{moreno2019toward}.

This paper introduces VitaNova -- an ALife simulation that investigates the evolution of neuronal agents, or ``boids,'' that form flocks to navigate and thrive in predator environments. This simulation revolves around two processes: self-organization, which is governed by evolving neural networks that dictate boid behaviour, and natural selection, which forces these agents to adapt and survive. This subtle interplay between individual behaviour modulation and group dynamics results in the formation of cohesive groups capable of collective reproduction—a phenomenon that mirrors key aspects of ETIs. VitaNova demonstrates how the combined forces of self-organization and natural selection can drive the spontaneous formation of reproductive groups, providing new insights into the evolution of complex biological organisation.

The next section presents results from VitaNova, which show the emergence of ring-structures that intriguingly exhibit the capacity for self-reproduction. We tackle the underlying question of how such complex collective behaviors arise. Through the simulation, we observe boids evolving their neural networks, thereby discovering increasingly adaptive behaviours that enable them to organize into flocks. This flocking behaviour serves as a survival mechanism, allowing them to more effectively evade predators and efficiently access resources. This evolutionary process, driven by natural selection at the individual level, eventually leads to the spontaneous formation of ring structures. These structures not only maintain stability over time but also demonstrate the ability to reproduce, forming new cohesive units. This phenomenon serves as a compelling indication of an ETI, showcasing the transition from individual boids to a higher-level reproductive entity within the simulation.

Following that, we provide a detailed description of the VitaNova model and how it differs from other ALife models having self-replicating patterns. The following sections delve into related work, providing a brief discussion of the vast topic of evolutionary transitions in individuality (ETIs) and situating our findings within this larger context. The paper concludes with an examination of potential future directions, laying out a road map for further research into the mechanisms and phenomena associated with ETIs as observed using the VitaNova simulation framework. The source code for the VitaNova simulation discussed in this paper is available at \url{https://github.com/namlehai90/Evolving-Neuronal-Flocks}.

\section{2. Results and Analysis}

Our simulation results reveal the spontaneous emergence of collective behaviours and structures within simulated flocks \footnote{A video demonstration of these emergent behaviors in the VitaNova simulation can be viewed here: \url{https://www.youtube.com/watch?v=0fqAUwj0fDk}.}.  Notably, we observed the formation of distinct ring structures, their growth, and eventual division, which serves as a foundation for discussing the implications of these patterns on evolutionary transitions in individuality (ETIs).

\subsection{2.1 Emergence of Collective Structures}
\label{sec:collective_structures}

The dynamics within VitaNova demonstrate the spontaneous emergence of complex collective structures from simple boid interactions. Significantly, the simulation shows a transition from solitary boids to organised flocks, culminating in the formation of ring structures that exhibit both stability and the capacity for division, indicating a potential mechanism for group-level reproduction.

\begin{figure}[h]
\centering
\includegraphics[width=0.35\textwidth]{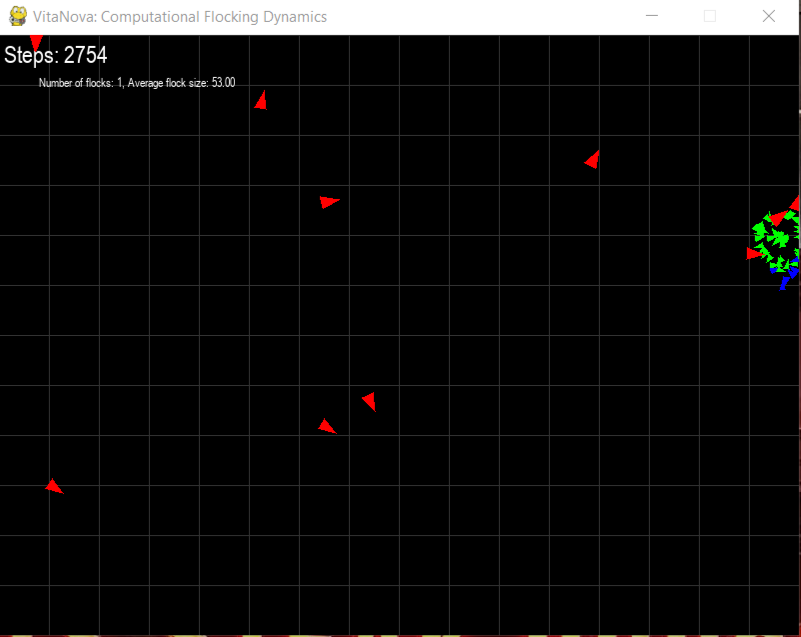}
\caption{\small Snapshot of VitaNova simulation showcasing a single stable ring structure prior to division. Boids are represented by green/blue triangles, forming a cohesive ring pattern, while predators are depicted as red triangles.}
\label{fig:pre_division}
\end{figure}

Figure \ref{fig:pre_division} captures a moment where the simulation exhibits a single, stable ring structure. This configuration remains consistent over numerous time steps, suggesting that the collective behaviour of boids leads to a self-organized equilibrium state, which persists despite environmental pressures and the presence of predators (red triangles).

\begin{figure}[h]
\centering
\includegraphics[width=0.35\textwidth]{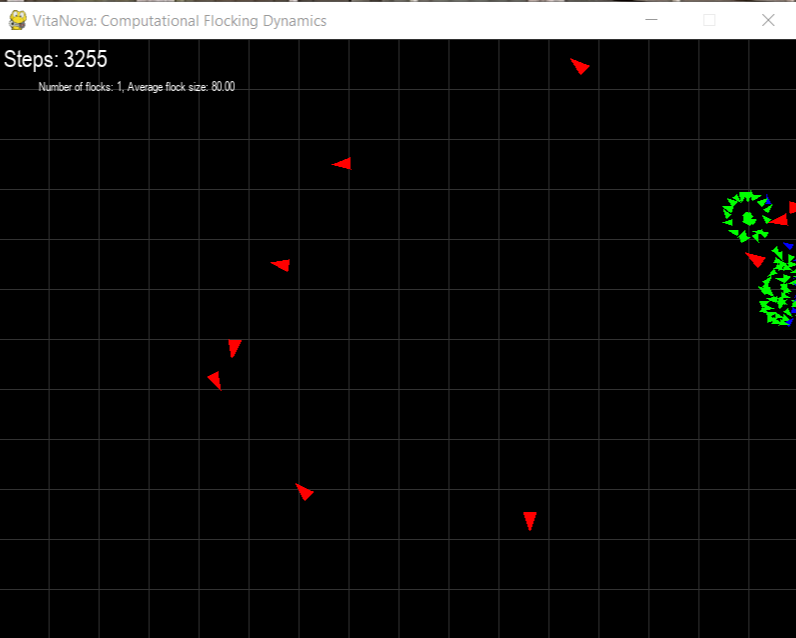}
\caption{\small Post-division state within VitaNova, where the initial ring has divided into two distinct flocks, embodying a form of collective reproduction.}
\label{fig:post_division}
\end{figure}

As the simulation progresses, we observe a critical transformation in the flock structure. Figure \ref{fig:post_division} shows the aftermath of a division event, where the initial ring has split into two separate entities. This bifurcation event is a hallmark of emergent reproductive behaviour at the group level, analogous to cell mitosis in biological systems \cite{mazia1961mitosis}.

\begin{figure}[h]
\centering
\includegraphics[width=0.35\textwidth]{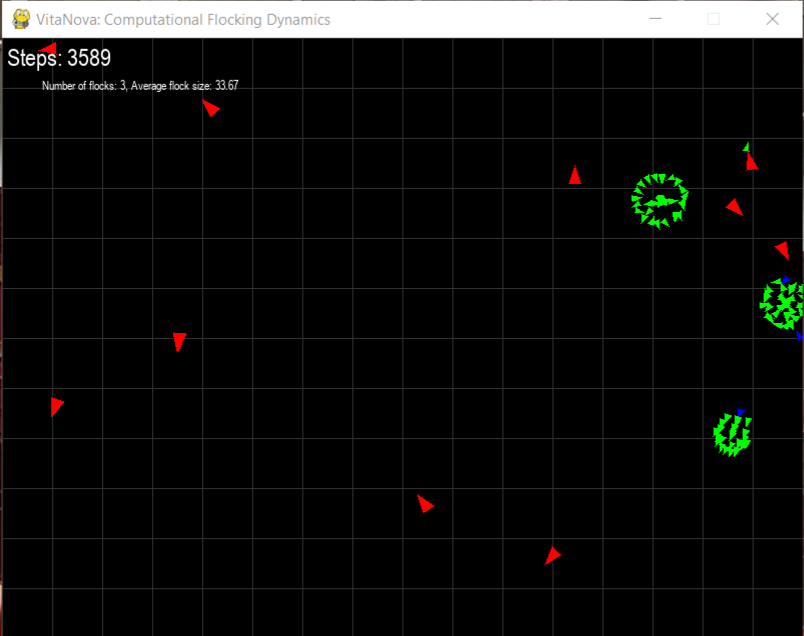}
\caption{\small The system after a second division event, displaying two distinct ring structures. This phase of the simulation suggests a cyclical pattern of growth and division, a characteristic of emergent collective reproduction.}
\label{fig:post_second_division}
\end{figure}

Lastly, Figure \ref{fig:post_second_division} captures the system after the second division, showing two distinct ring structures. This event indicates a cyclical pattern where reaching a certain size triggers division, reminiscent of cell division in biology. Notably, the second 'offspring' flock has drifted towards the bottom of the screen, distancing itself from its `parent' flock.

The succession of these division events reinforces the concept of emergent collective reproduction within our simulation. The flocks do not merely grow in numbers but undergo a division that hints at a form of reproduction at the collective level. Such behaviour emerges from the underlying rules governing individual boids, without the need for explicit programming of reproductive synchronisation. This cyclical pattern of collective growth and division provides a compelling demonstration of how complex life-like behaviours can spontaneously emerge from simple rules in artificial life systems.

\subsection{2.2 Group Reproduction in First-Generation Ring Structure}

\begin{figure}[h]
\centering
\includegraphics[width=0.35\textwidth]{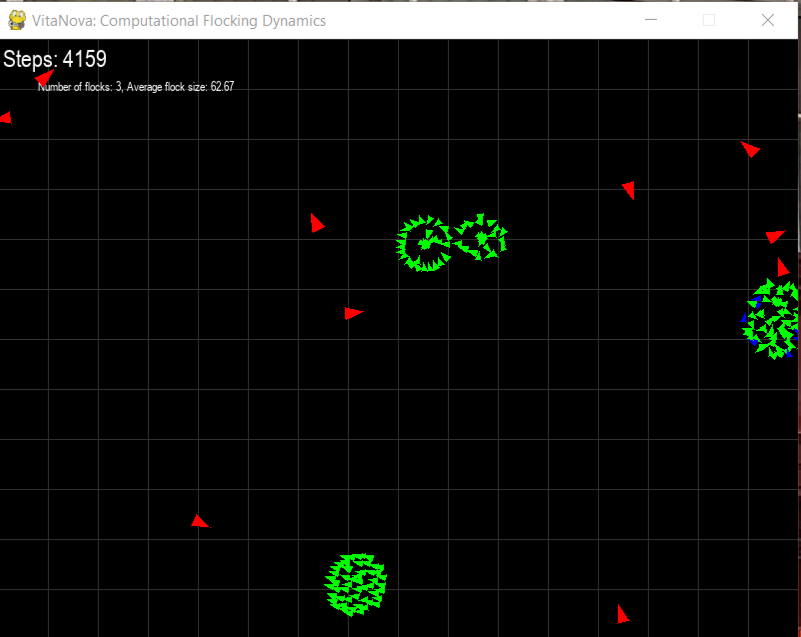}
\caption{\small The system post second division, showing growth in the average flock size and indicating a preparation for further reproductive events.}
\label{fig:pre_third_division}
\end{figure}

The development and subsequent reproduction of the first-generation ring structure is an interesting event in our simulation. Figure \ref{fig:pre_third_division} demonstrates the maturity of this ring structure as it prepares for its own reproductive event. The notable increase in its average size is indicative of the accumulation of resources and energy, setting the stage for the emergence of a new collective entity.

\begin{figure}[h]
\centering
\includegraphics[width=0.35\textwidth]{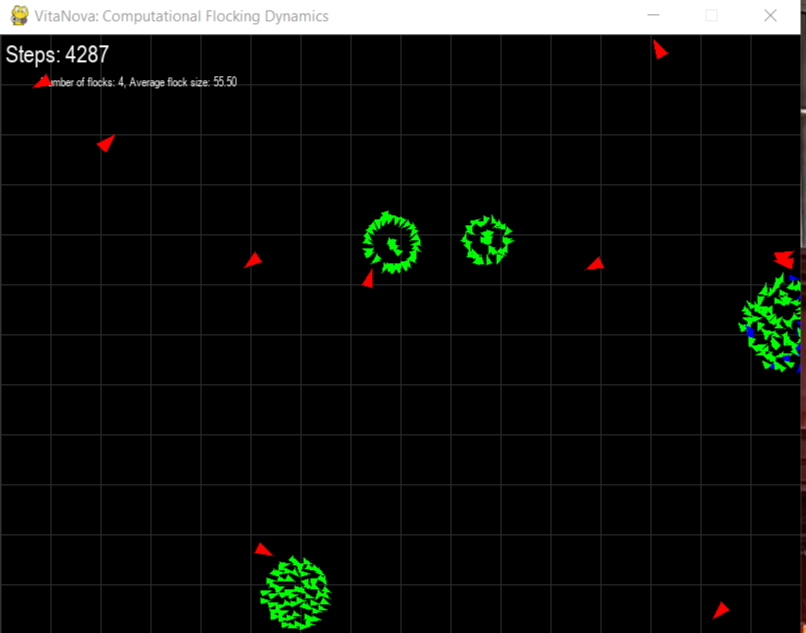}
\caption{\small The formation of new ring structures following the division of the first-generation ring offspring, showcasing the self-replicating nature of collective entities.}
\label{fig:post_third_division}
\end{figure}

Subsequently, Figure \ref{fig:post_third_division} captures a significant moment in the simulation—the first-generation ring offspring undergoing its own division, resulting in the creation of two new ring structures. This event underscores the simulation's capacity for illustrating cyclical patterns of collective reproduction, demonstrating the entities' ability for self-sustained replication.

The observation of these successive divisions, leading to self-similar structures, suggests a complex mimicry of biological reproduction processes within our artificial life model. This pattern of dynamic, repetitive division among ring structures hints at a core aspect of life: the ability to self-replicate. Remarkably, this capability emerges spontaneously from the simulation, devoid of explicit preprogramming for such intricate behaviours, offering profound insights into the mechanisms of emergent complexity in systems resembling life.

\subsection{2.3 Observational Analysis of Flock Metrics}

\begin{figure}[h]
\centering
\includegraphics[width=0.5\textwidth]{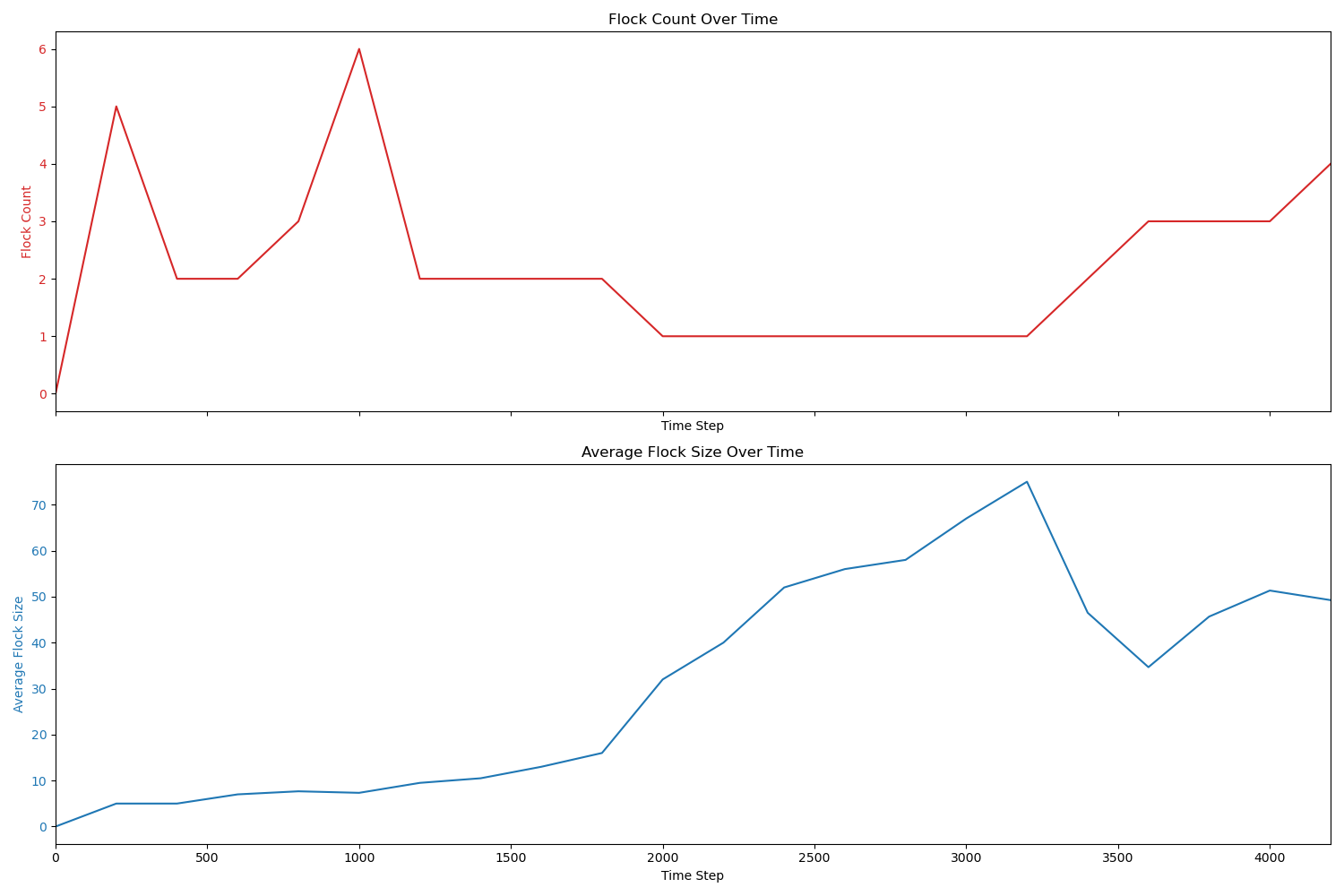}
\caption{\small Comparison of flock size and count over time.}
\label{fig:flockDynamics}
\end{figure}

The evolution of flock metrics over time offers a compelling narrative of the underlying dynamics governing the VitaNova simulation. Figure \ref{fig:flockDynamics} presents a dual perspective on the flock count and average flock size, which, when analysed together, indicate a periodicity suggestive of reproductive events.

An initial fluctuation in flock count reflects the turbulent phase of flock formation and disbandment as the system seeks stability. As the simulation progresses, we observe a period where a single flock dominates, aligning with a steady increase in average flock size (steps 2000 to 3200). This growth, uninterrupted by the formation of new flocks, suggests an accumulative phase where the collective structure fortifies, potentially gearing up for a reproductive division.

The eventual bifurcation into two flocks corresponds with a sharp decline in average flock size, connoting a successful reproductive event. The subsequent increase in average flock size within these new collectives suggests a repeat of the cycle, pointing towards an intrinsic mechanism that drives the collective towards replication once a threshold is crossed.

These cycles of growth and division, devoid of external control, underline the spontaneous emergence of group reproduction. They resonate with our visual observations, where flocks exhibit stable growth followed by a sudden division, mirroring processes observed in natural systems. Thus, our model offers a tangible representation of emergent collective reproduction, laying the groundwork for future studies into the complex behaviours that characterise evolutionary transitions in individuality.

Additionally, the increase in flock size prior to each division event signifies ongoing reproductive activities among the individual boid components within the collective. This continuous growth within the ring structures not only prepares the collective for division but also indicates that reproduction at the individual level contributes significantly to the overall process of collective replication.

\subsection{2.4 Role Distribution and Collective Adaptations}
\label{subsec:analysis_initial_flock_offspring}

This section delves into the adaptive dynamics of Flock 0 and its first-generation offspring, Flock 1, highlighting their distinct strategies in response to predator pressures and internal growth demands.

\begin{figure}[h]
\centering
\includegraphics[width=0.35\textwidth]{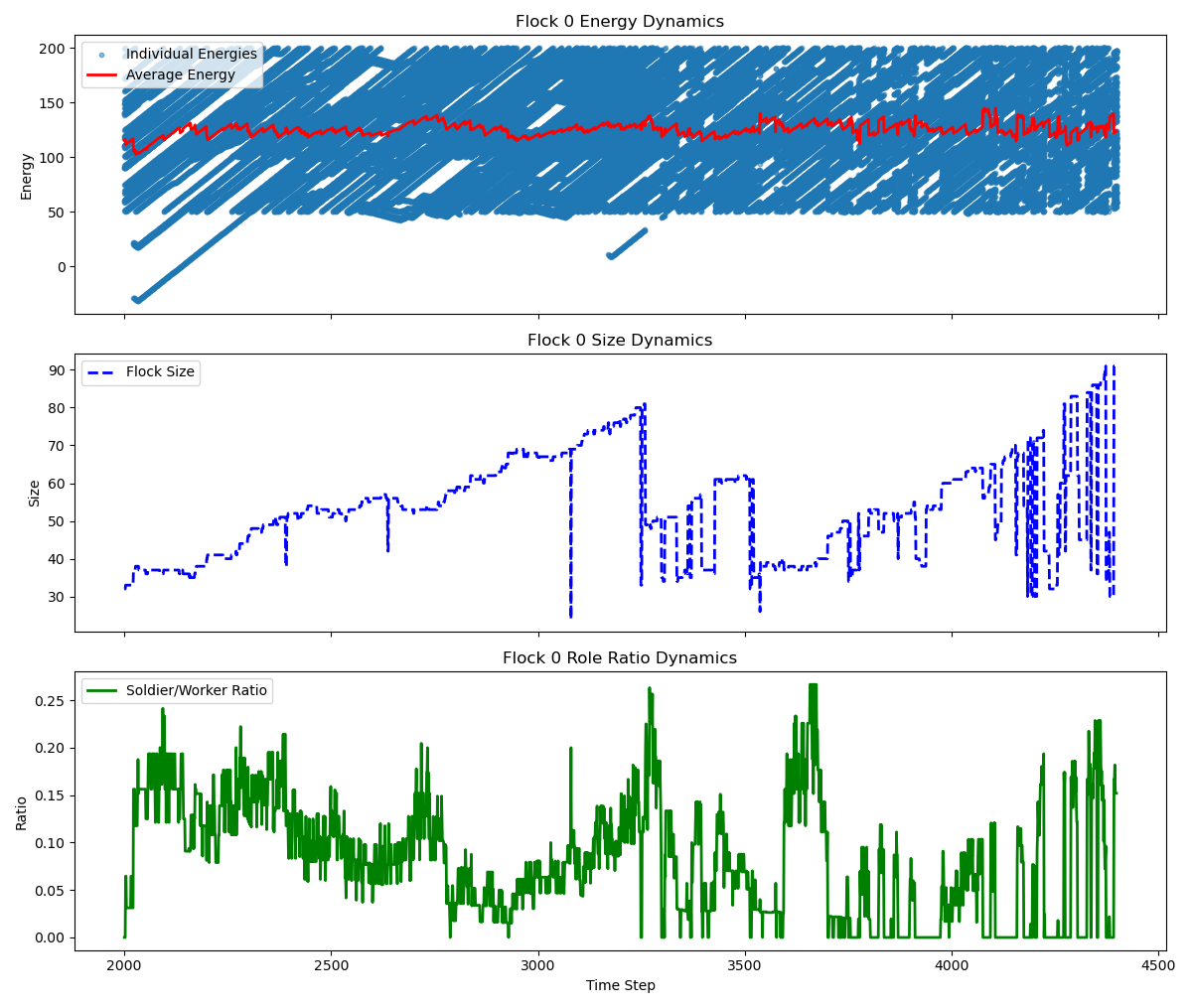}
\caption{\small Combined energy, size, and role ratio dynamics for Flock 0 over time.}
\label{fig:dynamics_flock0}
\end{figure}

Figure \ref{fig:dynamics_flock0} illustrates Flock 0's strategic equilibrium between energy distribution and flock size. Despite individual energy variances, a consistent average suggests a pre-reproductive balance. Interestingly, the soldier/worker ratio inversely correlates with flock size, indicating a tactical shift from defence to expansion in anticipation of reproduction (Figure \ref{fig:pre_division}). This shift underscores Flock 0's emergent strategy to balance survival and growth in a predator-dense environment.

\begin{figure}[h]
\centering
\includegraphics[width=0.35\textwidth]{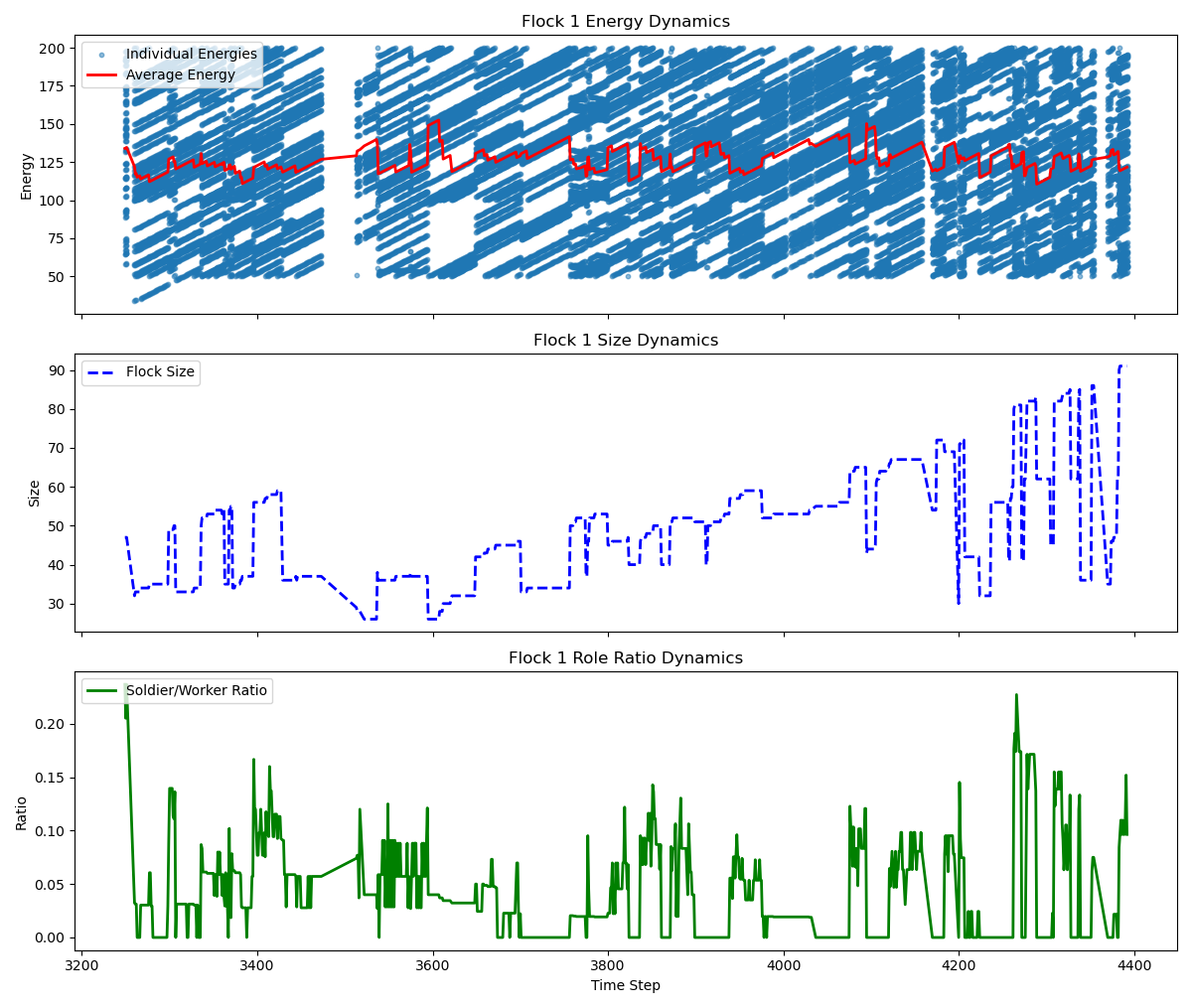}
\caption{\small Combined energy, size, and role ratio dynamics for Flock 1 over time.}
\label{fig:dynamics_flock1}
\end{figure}

Contrastingly, Flock 1 (Figure \ref{fig:dynamics_flock1}) showcases a proportional growth-defence strategy, diverging from its predecessor. Notably, Flock 1's adaptation in a less predator-threatened environment optimises its resource allocation towards balanced growth, maintaining a steady soldier/worker ratio. This adaptation indicates a sophisticated response, leveraging environmental conditions to strategise expansion while ensuring defence readiness.

The differential response to predator presence between Flock 0 and Flock 1 underscores the capacity of such artificial life systems to exhibit advanced forms of adaptive behaviour. It suggests that collectives within the simulation can not only evolve complex internal dynamics but also react and adapt to the broader environmental context in which they exist. This ability to modify strategies based on environmental pressures further affirms the potential of simulations like VitaNova to explore the intricate dynamics of life and evolution, offering insights into how entities might navigate and adapt to changing ecological landscapes

\section{3. Model Description}

The results presented arise from VitaNova, whose name translates to ``New Life'' from Latin. This platform models the interactions of autonomous agents, or boids, under the influence of environmental pressures such as predators and spatial constraints, to uncover patterns akin to evolutionary transitions in individuality (ETIs) without explicit outcome programming. \footnote{The implementation details of our models are fully available for review and use at \url{https://github.com/namlehai90/Evolving-Neuronal-Flocks}.
}

\begin{figure}[ht]
\centering
\includegraphics[width=0.35\textwidth]{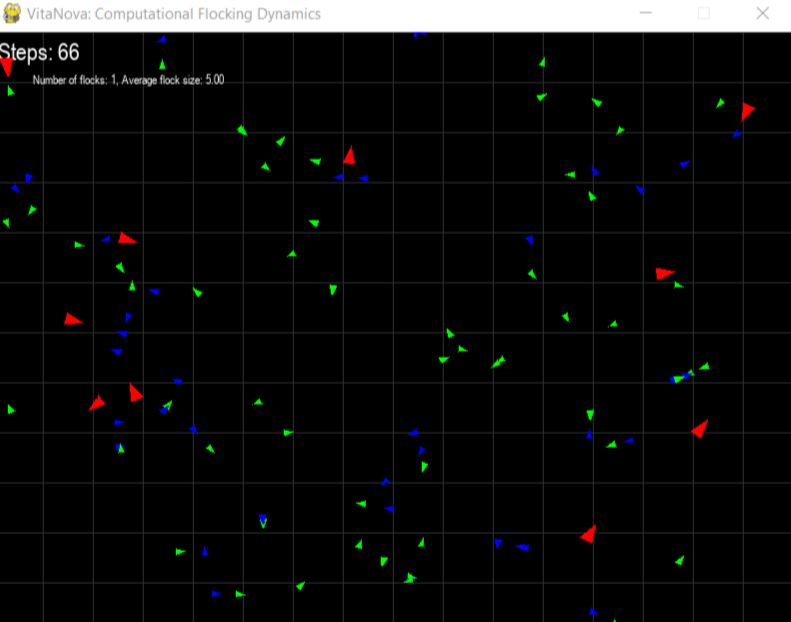}
\caption{\small Schematic of the VitaNova simulation environment, showing worker boids (green), soldier boids (blue), and predators (red triangles) within a 160 * 120 boid-unit 2-dimensional space. Key parameters include a BOID\_SIZE of 5 units and PREDATOR\_SIZE of 10 units, underscoring the spatial and interactive dynamics at play.}
\label{fig:environment}
\end{figure}

Notably, VitaNova abstracts the concept of resource consumption to simplify the model, omitting direct food gathering in favor of a focus on interaction-driven behavioural and evolutionary developments. The accompanying Figure \ref{fig:environment} visualises VitaNova's setup, depicting boids navigating a space defined by specific parameters, with predators introduced as a selective pressure on the population.

\subsection{3.1 VitaNova Agent Design and Neural Control}
\label{subsec:vitanova_agent_design}

The VitaNova simulation equips each boid with a neural network, enabling dynamic navigation through the environment. This approach departs from \cite{reynolds1987flocks}'s classical model  by utilising neural outputs to modulate core steering behaviours --- separation, alignment, and cohesion --- as opposed to fixed rules and constants.

\subsection{Neural Network Architecture}

\textbf{Input Layer:} The network's input layer comprises 11 neurons, processing a range of environmental cues such as the relative positions and velocities of nearby boids and predators, alongside the boid's current energy level. This setup allows for a detailed perception of the surrounding environment and the boid's status within it.

\textbf{Hidden Layer:} A singular hidden layer with 10 neurons intermediates the input signals, extracting relevant patterns to inform the boid's behavior.

\textbf{Output Layer:} The final layer outputs tendencies for each behavioral aspect:
\begin{itemize}
    \item \textbf{Separation Tendency} ($T_{sep}$): Adjusts the boid's effort to maintain a safe distance from neighbors, reducing overcrowding.
    \item \textbf{Alignment Tendency} ($T_{align}$): Influences the boid to synchronize its movement direction with that of nearby boids, enhancing group cohesion.
    \item \textbf{Cohesion Tendency} ($T_{coh}$): Directs the boid towards the mean position of nearby boids, fostering collective unity.
    \item \textbf{Predator Avoidance Tendency} ($T_{avoid}$): Initiates evasive actions when predators are close, improving survival chances.
    \item \textbf{Role Tendencies} ($T_{worker}$, $T_{soldier}$): Dictate the likelihood of the boid adopting worker or soldier roles in subsequent steps.
\end{itemize}

\begin{figure}[ht]
\centering
\includegraphics[width=\columnwidth]{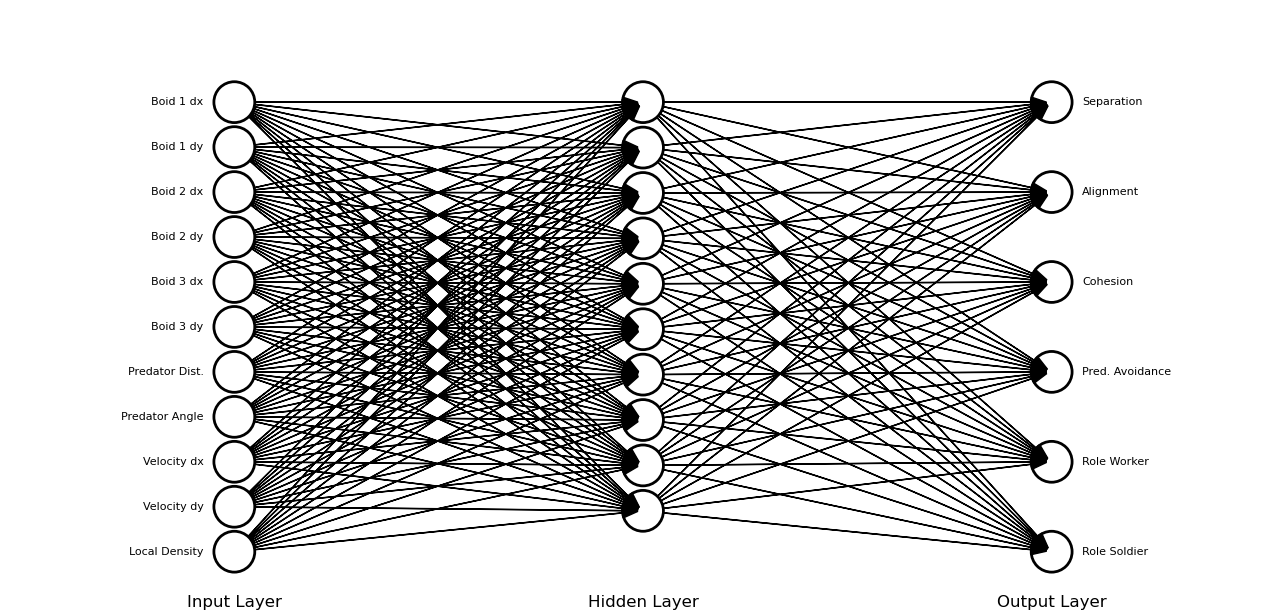}
\caption{\small Schematic representation of the neural network architecture controlling a boid in VitaNova. The input layer receives 11 distinct signals corresponding to the boid's sensory perception of its environment, including the relative positions and velocities of the closest neighbors (up to three), the nearest predator's position and velocity, and the boid's current energy level. These inputs are processed through a hidden layer consisting of 10 neurons, ultimately determining the boid's behavioral outputs. The outputs include tendencies for separation, alignment, cohesion, predator avoidance, and role selection (worker or soldier), which collectively dictate the boid's navigation and interaction within the simulation.}
\label{fig:neural_network_architecture}
\end{figure}

VitaNova leverages neural network-driven behaviour modulation to extend beyond the fixed behaviours of Reynolds' model, achieving a dynamic simulation environment. This allows boid behaviours to adapt in real-time to both environmental stimuli and their internal states, as detailed in Figure~\ref{fig:neural_network_architecture}. The neural architecture underpinning each boid enables a complex interplay between autonomy and collective behaviour, effectively capturing the nuanced dynamics of natural life forms. Thus, the neural network is crucial in VitaNova, facilitating an exploration of the interplay between individual actions and collective phenomena.

\subsubsection{Agent Motion and Steering}
\label{subsubsec:agent_motion_steering}

Boid motion within VitaNova is driven by neural network outputs, influencing core steering behaviors—separation, alignment, and cohesion—based on Reynolds' principles \cite{reynolds1987flocks}, with the addition of predator avoidance and role determination:

{\small
\begin{align*}
    \Delta(dx) =\ &T_{sep} \cdot Separation_x + T_{align} \cdot Alignment_x \\
                 +\ &T_{coh} \cdot Cohesion_x + T_{avoid} \cdot Avoidance_x, \\
    \Delta(dy) =\ &T_{sep} \cdot Separation_y + T_{align} \cdot Alignment_y \\
                 +\ &T_{coh} \cdot Cohesion_y + T_{avoid} \cdot Avoidance_y,
\end{align*}
}

where $\Delta(dx)$ and $\Delta(dy)$ denote adjustments to the boid's velocity in the x and y directions, respectively, and $T_{avoid}$ represents the predator avoidance tendency with $Avoidance_x$ and $Avoidance_y$ as the directional components of evasion.

Velocity is updated accordingly:
{\small
\begin{align*}
    dx_{new} &= dx + \Delta(dx), \\
    dy_{new} &= dy + \Delta(dy).
\end{align*}
}

Role-switching is determined by the outputs $T_{worker}$ and $T_{soldier}$, with the boid adopting the role with the higher tendency at each simulation step, facilitating dynamic adaptation to the environment and task demands.
    
The predator avoidance mechanism in VitaNova is governed by the neural output $T_{\text{avoid}}$, which dynamically modulates each boid's evasion behavior. The intensity of avoidance, determined by $T_{\text{avoid}}$, influences whether a boid opts for individual escape or seeks safety in numbers by flocking with others. This behavior is encapsulated by the following equations:

{\small
\begin{align*}
    (\text{escape\_dx}, \text{escape\_dy}) &= \text{escape\_predator}(\text{nearest\_predator}, T_{\text{avoid}}),
\end{align*}
}

where $\text{escape\_predator}(\cdot)$ computes the evasion vector $(\text{escape\_dx}, \text{escape\_dy})$ away from the closest predator, scaled by the avoidance intensity $T_{\text{avoid}}$. The boid's velocity is then updated as follows to incorporate this evasion vector:

{\small
\begin{align*}
    dx_{\text{new}} &= dx + \text{escape\_dx}, \\
    dy_{\text{new}} &= dy + \text{escape\_dy}.
\end{align*}
}

This critical neural output enables a boid to dynamically decide between an individual strategy to evade predators or to coalesce with other boids, potentially enhancing its survival through collective avoidance. The adaptability and complexity of these evasion tactics underscore the sophisticated interplay between autonomy and collective behavior in the VitaNova simulation, mirroring survival strategies observed in natural systems.

\subsection{3,2 Predator Behaviour}

In addition to boids, predators are introduced to provide dynamic challenges, influencing boid behaviours and flock evolution. With behaviorus defined by speed, visual range, and a catch radius, predators hunt within the simulation, reversing direction when encountering three or more soldier boids. This adaptive behaviour simulates predator aversion to collective defences, emphasising the soldier role's importance in boid survival strategies. Such predator-boid interactions enrich the ecosystem's complexity, showcasing the evolutionary pressures that shape group behaviours, social structures, and the adaptive role-switching between worker and soldier boids. This dynamic interplay enhances our understanding of natural selection and the development of cooperative strategies within artificial life simulations.

\subsection{3.3 Evolutionary Dynamics and Emergent Properties}

VitaNova's evolutionary mechanism is based on energy-driven reproduction and neural network mutations. When a boid accumulates sufficient energy, it reproduces, creating an offspring whose neural parameters are mutated versions of its own. This method facilitates behavioural evolution and strategy diversification within the population. Algorithm \ref{alg:reproduction} is the pseudocode for how evolutionary algorithm works in VitaNova. This approach ensures the continuous introduction of genetic variability, allowing the emergence of novel behaviours and enhancing the population's overall adaptability.

\begin{algorithm}
\small
\caption{Pseudocode for Reproduction and Mutation}
\begin{algorithmic}[1]
\For{\textbf{each} boid}
    \If{\texttt{boid.energy} $>$ \texttt{REPRODUCTION\_THRESHOLD}}
        \State \texttt{boid.energy} $-=$ \texttt{REPRODUCTION\_COST}
        \State offspring $\gets$ \texttt{create\_offspring(boid)}
        \For{\textbf{each} weight \textbf{in} offspring.neural\_network.weights}
            \If{\texttt{random()} $<$ \texttt{MUTATION\_RATE}}
                \State weight $+=$ \texttt{normal(0, MUTATION\_STRENGTH)}
            \EndIf
        \EndFor
    \EndIf
\EndFor
\label{alg:reproduction}
\end{algorithmic}
\end{algorithm}

In VitaNova, fitness emerges implicitly rather than being dictated by an external objective function typical of traditional evolutionary algorithms for problem-solving \cite{fogel1994introduction, mitchell1998introduction}. Unlike these models, where behavioural outcomes might be somewhat predefined or directed, VitaNova's interesting property lies in its unguided evolution of behavioural complexity, aligning more closely with natural selection in the biological world. This mirrors approaches in artificial life such as Tierra by Thomas Ray and PolyWorld by Larry Yaeger, where complex behaviors and structures arise from the interactions within the system \cite{ray1993evolutionary, yaeger1994computational}. This lack of explicit fitness criteria allows for the spontaneous emergence of organisational structures and survival strategies, underscoring VitaNova's potential to explore the foundations of adaptive systems and complexity evolution without human-imposed direction.

\subsection{3.4 Distinction from Other ALife Models}

VitaNova distinguishes itself in the artificial life (ALife) domain by its integration of self-organization and natural selection at the level of individual components, setting it apart from classical CA-based model like Conway's Game of Life \cite{bak1989self}, and Langton's Loop \cite{langton1984self}, and the more recent Lenia \cite{chan2020lenia}. While these models excel in showcasing complex replicating patterns and complex behaviours emerging from simple rules through self-organization, they generally lack mechanisms for simulating evolution via natural selection for individual entities.

\textbf{Fusion of selection and self-organization in VitaNova:} Central to VitaNova's design is the simultaneous operation of self-organization and natural selection processes. This dual mechanism ensures that while boids autonomously generate complex patterns and structures through interaction, they also evolve through natural selection based on their fitness—determined by their ability to navigate environmental challenges and reproduce. This results in a dynamic evolutionary landscape where adaptations are continually refined in response to shifting environmental pressures.

VitaNova exemplifies the convergence of self-organization and natural selection, embodying a debate central to evolutionary theory. This platform demonstrates how individual boids, through interactions and fitness-driven natural selection, navigate and adapt to environmental pressures, leading to complex, emergent behaviours and evolutionary advancement. This synthesis addresses classical discussions, such as those between Stuart Kauffman, who emphasises self-organization's role in biological complexity, and John Maynard Smith, who underscores the primacy of natural selection \cite{kauffman1993origins, smith1993theory}. Additionally, it resonates with more recent perspectives on the interplay between these forces in evolutionary processes, such as  \cite{weber1996natural, batten2008visions}. VitaNova, by integrating these mechanisms at the individual level, offers insights into the ongoing debate about the origins of complex adaptive systems and the emergence of higher organizational levels in the natural world.

\section{4. Further Discussion and Related Work}

The quest to understand Evolutionary Transitions in Individuality (ETIs) has engaged scholars across disciplines, leading to a wealth of theoretical frameworks aimed at elucidating the complex processes that drive the emergence of higher-order biological organisations from simpler life forms. Seminal works by John Maynard Smith and Eörs Szathmáry have proposed models that describe key stages and mechanisms in the hierarchical complexity evolution of life \cite{szathmary2015toward}. Building on these foundations, recent contributions have expanded the discourse, introducing diverse perspectives that challenge and enrich our understanding of ETIs \cite{szathmary2015toward, godfrey2011darwinian}. Among these, Watson, Levin, and Buckley offer a comprehensive summary of current debates, highlighting viewpoints such as the challenge of defining levels of selection and individuality \cite{okasha2005multilevel}, the role of genetic relatedness \cite{west2015major} \cite{watson2022design}. A contentious debate within the field revolves around the necessity and timing of higher-level selection mechanisms. Some theories propose that higher levels of individuality emerge through selection pressures acting on group traits, raising questions about the pre-existence of such selection mechanisms or their concurrent development with the transition itself. In addressing these complex themes, Watson et al. lean towards an explanation of ETIs that heavily favors the self-organization process, positing a less emphasized role for natural selection. They propose an intriguing parallel between evolutionary transitions and neural network learning, suggesting that the process of evolution by natural selection and the organization within neural networks might follow similar patterns of emergent behaviors and hierarchical organization from simple units \cite{watson2022design}.

These models, while provide solid foundations, often remain abstract, exploring ETIs through mathematical and philosophical lenses without providing empirical demonstrations of the underlying processes \cite{watson2022design, west2015major, michod2003reorganization}. A notable challenge within this theoretical exploration is the simulation of group reproduction — a critical marker of ETIs — which many models assume rather than demonstrate. This gap highlights a need for dynamic models that can capture the emergent behaviors and complex interactions leading to the spontaneous formation of cooperative, reproductive units \cite{hanschen2015evolutionary, black2020ecological}.

Transitioning from theoretical models to Artificial Life (ALife) simulations offers a promising avenue for addressing these limitations. ALife's potential in modelling complex biological phenomena has been recognized, yet realizing simulations that accurately reflect the dynamics of ETIs poses its own set of challenges. John E. Stewart, in his discussions on ALife's role in studying ETIs, emphasizes the difficulty of simulating these transitions without resorting to hierarchical control mechanisms akin to "vertical bosses" observed in natural systems like insect colonies. Stewart advocates for models that can account for the emergence of cooperative behavior and higher-order individuality through self-organizing processes, underscoring the importance of managerial roles in facilitating ETIs \cite{stewart1997evolutionary, stewart2020towards}.

VitaNova diverges from these perspectives by integrating both self-organization and natural selection to demonstrate ETIs, even without the ``vertical boss''. Contrary to models that might prioritise self-organization processes with minimal engagement of selection, VitaNova showcases that the spontaneous emergence of collective reproduction, emblematic of ETIs, can result from the intricate interplay between evolving neuronal agents' self-organized behaviors and the selective pressures they encounter. This balanced approach not only highlights group reproduction but also serves as a dynamic model for investigating the conditions under which such higher-order biological organisation emerges, thus providing empirical insights into the spontaneous formation of cooperative, reproductive units.

In recent developments within artificial life research, the work of \cite{moreno2019toward} on the DISHTINY platform presents a crucial step towards understanding fraternal transitions in individuality. Their simulation models cell-like entities that demonstrate cooperative behaviours, reproductive division of labour, and hierarchical emergence of individuality, offering insights into the conditions fostering complex organismal structures. However, our VitaNova simulation diverges in its approach by emphasising the spontaneous emergence of collective reproduction without pre-defined cooperative strategies. Our simulation leverages neural network-driven autonomous agents, capable of evolving adaptive behaviours through interactions with the environment and each other. Unlike DISHTINY, where spatial and temporal coordination is designed to maximise resource harvest, VitaNova's agents self-organize into reproducing groups based on internal dynamics and natural selection, without direct rewards for cooperation or aggregation. This fundamental difference highlights VitaNova's capacity to explore the open-ended evolution of behavioural complexity and the conditions under which higher levels of biological organisation can emerge from lower level entities.

\section{5. Conclusion, and Future Avenues}

This research elucidates ETIs by showcasing, via simulation, the emergence of reproductive groups from the synergy of self-organization and natural selection. Neuronal agents within our simulation autonomously evolve adaptive rules, leading to self-organized flocking behaviours. These flocks then act as unified entities capable of reproduction, highlighting a potential pathway for the emergence of new levels of biological individuality. Post-simulation analyses reinforce these observations, offering empirical support to the theoretical framework posited by our simulation outcomes.

Future research will focus on expanding the simulation's capabilities and enhancing our analytical tools. Key directions include:

\begin{itemize}

    \item Introduce more complex genetic and environmental dynamics, including evolving predator behaviours, seasonal changes, and resource variability. This expansion aims to simulate an evolutionary arms race between predators and boids, providing insight into its impact on the emergence of higher-level organisational structures and offering a more detailed exploration of ETI phenomena.
    
    \item Delve into the emergence of reproductive division of labour within groups to understand how specialised roles affect group fitness and social structure. This investigation aims to elucidate the dynamics and benefits of such specialisation, particularly in enhancing the adaptability and cohesion of reproductive collectives.
    
    \item Create a comprehensive framework for analysing emergent behaviours, incorporating metrics to evaluate cooperation, conflict resolution, and resource allocation efficiency within groups. This approach aims to quantify the adaptive and organisational complexities of emergent reproductive collectives, facilitating a deeper understanding of their evolutionary success.
    
    \item Implement measures of genetic diversity and develop quantitative methods for assessing group selection. This initiative seeks to empirically test multilevel selection theory hypotheses in ETIs \cite{okasha2005multilevel}, shedding light on the evolutionary mechanisms that drive the formation of cohesive, reproducing groups from individual entities.
    
    \item Assess entropy production changes within flocks around ETI events to explore the thermodynamics of system organisation and complexity increase. This analysis will provide insight into the energetic trade-offs inherent in the formation of higher-level biological structures, reflecting the principles of nonequilibrium thermodynamics in living systems \cite{ prigogine1977self}. By quantifying entropy variations, we aim to elucidate the cost-benefit dynamics of emergent group behaviours and reproductive strategies from a thermodynamic perspective, offering a novel angle on the evolutionary transitions into higher level organisms.
    
\end{itemize}

\section{Acknowledgements}
The authors gratefully acknowledge the support of grants 62230 \& 62220 (RW) and 62212 (ML), from the John Templeton Foundation, and BBRSC grant BB/P022197/1 (CB). \footnote{Further details and all computational resources used in this study are available online.\footnote{All relevant source code can be accessed at \url{https://github.com/namlehai90/Evolving-Neuronal-Flocks}.}
}

\footnotesize
\bibliographystyle{apalike}

\end{document}